\documentstyle[aps,prb,epsf]{revtex}
\begin{document}
\parindent=0pt
\def \be{\begin{equation}}
\def \ee{\end{equation}}
\def \bea{\begin{eqnarray}}
\def \eea{\end{eqnarray}}
\def \bn{{\bf n}}
\def \br{{\bf r}}
\def \bx{{\bf x}}
\def \cH{{\cal H }}
\def \cA{{\cal A }}
\def \cL{{\cal L }}
\def \cB{{\cal B }}
\def \cP{{\cal P }}
\def \cS{{\cal S }}
\def \bA{{\bf A}}
\def \bL{{\bf L}}
\def \bhn{{\bf \hat n}}
\def \half{{1\over 2}}
\def \bk{{\bf k}}
\def \bq{{\bf q}}
\def \nd{{\vphantom{\dagger}}}
\def \cP{{\cal P}}
\def \cO{{\cal O}}
\def \yBCO6{{ YBa$_2$\-Cu$_3$\-O$_{7-\delta}$ }}
\def \yBCO6{{ YBa$_2$\-Cu$_3$\-O$_{6.6}$ }}
\def \yBCO6x{{YBa$_2$\-Cu$_3$\-O$_{6+x}$}}
\def \yBCFO{{YBa$_2$Cu\-$_{2.55}$\-Fe$_{0.45}$\-O$_{y}$}}
\twocolumn[\hsize\textwidth\columnwidth\hsize\csname@twocolumnfalse\endcsname
\title{Andreev Peaks and  Massive Magnons in Cuprate SNS junctions}
\author{Assa Auerbach and Ehud Altman}
\address{ Department of
Physics, Technion, Haifa 32000, Israel.} \date{\today}
\maketitle
\begin{abstract}
The
projected SO(5) theory (pSO(5)) is used to resolve the puzzle of two distinct
energy gaps in high T$_c$ Superconductor-Normal-Superconductor  
junctions. Counter to conventional   theory of multiple Andreev reflections
(MAR), the differential resistance peaks  are associated with
the antiferromagnetic resonance observed  in neutron scattering,  and not with
Cooper pair breaking.  The pSO(5) and MAR theories  differ by the
expected  tunneling charges at the peaks. We propose that
shot noise experiments could  discriminate against the conventional 
interpretation.  
\end{abstract} 
\vskip2pc] 
\narrowtext
In  current transport through
high $T_c$ superconductor junctions, there seem to be  
{\em two} energy scales\cite{Deutscher}. The upper energy  is seen in tunneling
conductance\cite{tunn-exp}, and is identified with the ``pseudogap''
$\Delta_p$ which  appears  in magnetic resonance\cite{nmr} and
photoemmission\cite{arpes}. A lower gap, which scales differently
with hole doping,  manifests as peaks in the differential resistance of
Superconducting-Normal-Superconducting (SNS) Josephson junctions\cite{nesher}.
These peaks  have been interpreted using the conventional theory of
multiple Andreev reflections, following  Klapwijk, Blonder,
and Tinkham (KBT)\cite{kbt}.

\begin{figure}[htbp]
  \begin{center}
    \leavevmode \epsfxsize0.75\columnwidth
    \epsfbox{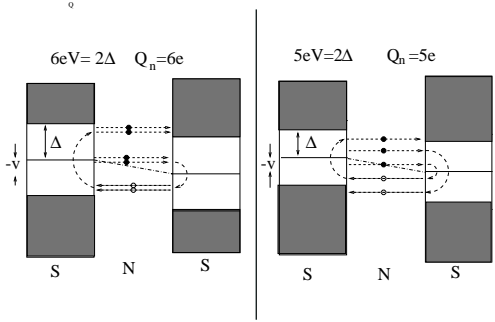}\vskip0.2pc
    \caption{{\bf KBT Theory.} Differential resistance peaks of  $n=6$  (left
diagram), and $n=5$ (right diagram), involve  a cascade of $n$
Andreev reflected charges traversing the normal metal. Singular dissipation is
due to emission of  quasiparticles above the $s$-wave gap.
Filled (empty) circles denote electrons (holes) in the normal barrier.}
\label{fig:btk} \end{center}\end{figure}

KBT theory treats two conventional superconductors
with a single $s$-wave BCS quasiparticle gap  $\Delta$,  separated by a free
electron metal. Electrons traversing the  metal are Andreev reflected back as
holes, gaining energy increments $eV$ at each traversal (as depicted in Fig.
\ref{fig:btk}).  Peaks in the differential resistance appear at voltages
$2\Delta/ne$, and are due to the $(E-\Delta)^{-1/2}$ singularity in the
quasiparticles'  density of states.
%%%%%%%%%%%%%%%%%%%%%%%%%%%%%%%%%%%
%%%%%%%%%%%%%%%%%%%%%%%%%%%%%%%%%%%
However, in cuprate SNS junctions, such as YBa$_2$\-Cu$_3$\-O$_{6.6}$  -
YBa$_2$Cu\-$_{2.55}$\-Fe$_{0.45}$\-O$_{y}$ -YBa$_2$\-Cu$_3$\-O$_{6.6}$
examined by Nesher and Koren\cite{nesher} , application of KBT theory is
problematic.   A naive  fit to KBT expression faces the two gaps puzzle, i.e.
an ``Andreev  gap'' is of order  $\Delta\approx 16$meV, while the
tunneling gap is about three times larger\cite{Racah}, and scales differently
with $T_c$.  Without perfect alignment of the interfaces, it is
hard to understand the observed sharpness of peaks\cite{nesher} since the
$d$-wave  gap is modulated for different directions. Moreover,  the barrier is
by no means a ``normal'' metal devoid of interactions:  it is
 an underdoped cuprate with  antiferromagnetic correlations and
strong pairing interactions as evidenced by a large proximity
effect\cite{prox}.

The purpose of this Letter is to provide an alternative explanation for
the differential resistance peaks series\cite{nesher}, which takes into
account the strong correlations in the pseudogap regime. Our analysis  resolves
the two energy scales puzzle.

We employ  the {\em projected} SO(5) (pSO(5)) model\cite{pSO5}, which
is a strong
coupling effective Hamiltonian. It  describes the dynamics and
interactions of four primary bosonic modes of  cuprates:   {\em preformed
hole pairs} and {\em massive spin one magnons}.

A  differential resistance peaks series is found at bias voltages
$V_n=\Delta_s /(en)$,
$n=1,2,\ldots$ where $\Delta_s$ is the
antiferromagnetic resonance energy. This resonance  has been directly
measured  by inelastic neutron scattering.   The peaks  are thus associated
with emmission of magnon pairs at the resonance threshold, and {\em not} with 
pair breaking, as in KBT theory. We note that other predictions to observe
magnons (also called $\pi$-modes) in various cuprate  junctions were
made\cite{SO5-JJ,AC},  but await experimental confirmation.
We propose that measurement of the excess
shot noise below the peaks,  could  discriminate against the latter
interpretation. pSO(5) theory predicts  tunneling charge
$2ne$ below the $n$-th peak,  while KBT theory expects charge  $ne$.

\vspace{0.5cm}
{\em Degrees of freedom}: At energies below the pseudogap $\Delta_p$, preformed
hole pairs  (with internal $d$-wave symmetry), describe the primary charge
degrees of freedom in the underdoped  regime\cite{uemura}. The hole pairs are
bosons, and their phase fluctuations are controlled by the two dimensional
superconducting stiffness $\rho_c$, as measured by the London penetration
depth. At $T_c$, the pairs Bose condense and long range phase coherence is
established. This scenario can explain\cite{phase-fluc}  the empirical
relations $T_c\propto \rho_c$, which have been observed in
cuprates\cite{uemura} at low doping concentrations.  The  other low
energy charge excitations are  fermionic quasiparticles near the  $d$-wave
nodes. These have a smooth density of states which decreases below
$\Delta_p$.

Additional  bosonic excitations
below the pseudogap energy scale,  are antiferromagnetic spin fluctuations
i.e.  magnons.  Massive spin one magnons  have been observed in inelastic
neutron scattering\cite{Fong} in YBa$_2$\-Cu$_3$\-O$_{6+\delta}$.  They
manifest as a sharp resonance in the spin correlation function
$S_{\alpha\alpha'}$, which near the antiferromagnetic wavevector $\bq\approx
\vec{\pi}$ has the form \be
S_{\alpha\alpha'}(\omega,\bq)\approx s_0
{\delta_{\alpha\alpha'} \over \omega^2-c^2(\bq-\vec{\pi})^2-\Delta_s^2 }
\label{Sqom}
\ee
Here $c$  is the spin wave velocity, and $s_0$ is a normalization factor.
The doping dependent resonance energy $\Delta_s(\delta) $ increases\cite{Fong}
between $\Delta_s(0.5)= 25$meV, (with $ T_c=52^\circ K$),
and  $\Delta_s(1)= 40$meV, (at $T_c=92^\circ K$)\cite{Fong}.
\vspace{0.5cm}
{\em The projected SO(5)  theory.} The large onsite Hubbard repulsion between
electrons is  imposed by an apriori projection  of doubly occupied  states
from the Hilbert space\cite{comm-SO5}.

 The undoped vacuum 
$|0\rangle$ is a half filled Mott insulator in a  quantum spin liquid state. 
The pSO(5) vacuum possesses short range antiferromagnetic correlations.
 A translationally invariant realization of $|0\rangle$ on the microscopic
square lattice, is the short range resonating valence bonds
state\cite{rvb,moshe}.

Out of this undoped
vacuum, $b^\dagger_{h}$ create charge $2e$ bosons (hole pairs) with internal
$d$-wave symmetry under rotations, and $b^\dagger_{\alpha}$, $\alpha=x,y,z$
create a triplet of antiferromagnetic,  spin one  magnons.

 The  lattice pSO(5) Hamiltonian is
\bea
\cH^{pSO(5)}&=&\cH^{charge} + \cH^{spin}+\cH^{int} +\cH^{Coul}+\cH^{ferm}
\nonumber\\
\cH^{charge} &=& (\epsilon_c- 2\mu) \sum_i b^\dagger_{h i } b^\nd_{ h i
} - {J_c\over 2} \sum_{\langle ij\rangle} \left( b^\dagger_{ h i } b^\nd_{ h i
}+\mbox{h.c.}\right)\nonumber\\ \cH^{spin} &=& \epsilon_s \sum_{i\alpha  }
b^\dagger_{ \alpha i} b^\nd_{ \alpha i} - {J_s} \sum_{\alpha \langle
ij\rangle} n^\alpha_i n^\alpha_j
\nonumber\\  \cH^{int} &=& W \sum_i
:\left( b^\dagger_{hi}b^\nd_{hi} +\sum_\alpha
b^\dagger_{ \alpha i}b^\nd_{ \alpha i} \right)^2 : , \nonumber\\
\label{pSO5}
 \eea
where $:():$ denotes  normal ordering, and
$n^\alpha_i=
(b^\dagger_{i\alpha}+b^\nd_{i\alpha})/\sqrt{2}$ is the  N\'eel spin field.
 $\cH^{int} $ describes  short range interactions between
bosons, and $\cH^{Coul}$
describes  the long range Coulomb interactions.  $H^{ferm}$ describes coupling
to the nodal (fermionic) quasiparticles, which  contribute to a large, but
smooth, conductance background. Here we will concentrate on the conductance
singularities, and will not compute the  fermionic  background.

The mean field approximation to
Eq. (\ref{pSO5}) is straightforward\cite{pSO5}. It amounts to replacing
$b^\dagger_{\gamma  i} \to \langle b^\dagger_{\gamma i}\rangle$,
$\gamma=h,\alpha $, and minimizing $\cH^{charge}+\cH^{spin}+\cH^{int}$  with
respect to     $\langle  b^\dagger_{\gamma i}\rangle$.   There is a first
order transition between the  two primary mean field phases on the square
lattice at $\mu=\mu_c$, where
\be  \mu_c={1\over 2}
(\epsilon_c-\epsilon_s)-(J_c-2J_s )  ,
\ee
$\mu_c$ is  of the order of the Hubbard interaction scale.

At $\mu<\mu_c$ we have an  undoped Mott insulator  
with  no hole
pair bosons, and where the magnons Bose-condense. The condensate  supports  a
finite staggered magnetization 
\be
 |\langle
n^\alpha\rangle|^2 = (2J_s-{1\over 2}\epsilon_s)/W   ~~~~~ \mu < \mu_c 
\ee
There are two linear 
spin wave modes $\omega=c|\bq|$, with where $c= \sqrt{2} J_s/\hbar $ is
the semiclassical spinwave  
velocity of the  Heisenberg antiferromagnet.

At  $\mu>\mu_c$ the ground state  becomes doped with hole pairs which
Bose-condense into a
superconducting phase with an order parameter
\be
|\langle b^\dagger_{i } \rangle|^2 = (\mu-\mu_c+2J_s-\epsilon_s/2)/W ~~~~~
\mu > \mu_c \label{Delta}
\ee
Long range interactions in $\cH^{Coul}$, 
frustrate  the first order transition and create
intermediate (possibly incommensurate) phases\cite{pSO5}, which we shall not
discuss here.

The mean field phase stiffness  is given by $\rho_c=J_c  \langle
b^\dagger_{i } \rangle^2$,   and therefore Eq. (\ref{Delta})  explains 
why $\rho_c$  increases with chemical potential (and doping) in the
underdoped superconducting regime, as observed  experimentally\cite{uemura}.

Analysis of the linear quantum fluctuations about mean field theory 
determines the magnon dispersion i.e. the  poles of  Eq. (\ref{Sqom}).
The mean field magnon gap is found to  be  
\be
\Delta_s=2\sqrt{(\mu-\mu_c)(\mu-\mu_c  +4J_s)}
\ee
which by Eq. (\ref{Delta}) implies that $\Delta_s^2 \propto \rho_c, T_c$.
Thus the pSO(5) mean field theory can explain the   systematic   increase
of $\Delta_s$ with $T_c$ which is
observed  by Fong {\it et. al.}\cite{Fong}.

\vspace{0.5cm}
{\em The cuprate SNS junction.}
We consider a  junction, where  the barrier (N)  has no
superconducting  or magnetic order $\langle b^\dagger_h\rangle=0,\langle
n^\alpha\rangle=0$.  We  derive on general grounds  the form of the
effective tunneling Hamiltonian between superconductors as follows.
An integration of the barrier's charged bosons  $b_{h}$ out of the path
integral results in an effective action $\cS^{tun}$ which couples the charges
of the two  superconductors . $\cS^{tun}[b_{h_L},b_{h_R},b_\alpha]$
explicitly  depends on the hole pairs  bosons on  the left and right
interfaces, and on the magnons  in the barrier. By charge conservation,  an
expansion of $\cS^{tun}$  as a power series leaves only terms with equal
number of $b_h$'s and $b^\dagger_h$'s. By spin conservation, the magnon
terms are singlets and hence at least bilinear in 
$n^\alpha$.
%%%%%%%%%%%%%%%%%%%%%%%%%%%%%%%%%%%
\begin{figure}[htbp]   \begin{center}
    \leavevmode \epsfxsize0.75\columnwidth
    \epsfbox{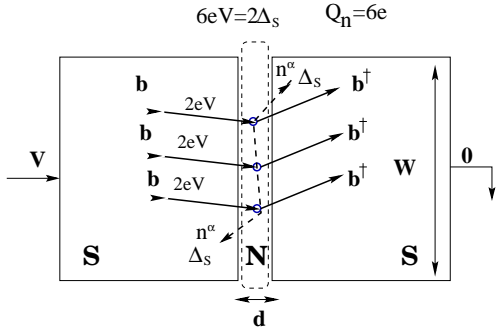}\vskip0.2pc
\caption{{\bf pSO(5) theory for Andreev peaks in cuprate SNS junctions}.
Three hole pairs co-tunneling from left to right, generate a pair of magnons.
At the antiferromagnetic resonance threshold $6eV=2\Delta_s$, this
process contributes to the $n=3$ peak of the differential resistance. The
diagram contains lowest order contributions of  hole pairs-magnon interactions
to the tunneling vertex $T_3$. } \label{fig:diag} \end{center} \end{figure}
%%%%%%%%%%%%%%%%%%%%%%%%%%%%%%%%%%%

This expansion leads to a series of tunneling terms. For the Andreev peaks
we  retain  only the leading order terms (in $b^\dagger,b$) which are
\bea
 \cH^{tun-mag} &=& -
\sum_{n} (\cA^\nd_n + \cA^\dagger_n) \nonumber\\ \cA_n&=& \sum_{y_1\ldots
y_{2n},\bx,\bx'} T_n  b^\dagger_{h_L, 1} \ldots b^\dagger_{h_L,n} ~b^\nd_{h_R
n+1} \ldots b^\nd_{h_R ,2n } \nonumber\\ && \times \left( \sum_\alpha
n^\alpha(\bx) n^\alpha (\bx') \right) \label{H-tun}
\eea
$\cA_n^\dagger$  describes  a simultaneous tunneling of $n$ hole
pairs from the left to the right superconductor,  coupled to
a magnon pair  excitation.  $T_n$ is the tunneling vertex function,
which depends on the bosons positions.  

 The energy transfer mechanism is depicted
diagrammatically in Fig.\ref{fig:diag}. We do not compute $T_n$'s
which depend on the details of the barrier and the interfaces.
A ``good'' N barrier is defined to have  sizeable $T_n$, if multiple pair
tunneling terms are to be observed. This requires a thin  barrier with slowly
decaying spin and  charge correlations\cite{prox}.
It is important to
note that multiple pair tunneling, i.e. the differential resistance peaks at  $
n>1 $, depends on strong   anharmonic  interactions between the hole pairs
and  magnons. {\em These  interactions are an  essential part of the pSO(5)
theory} as   modelled by $\cH^{int}$ in Eq. (\ref{pSO5}).

The junction's conductance is calculated
in the standard fashion\cite{mahan}: the bias voltage $V$
transforms  the left bosons $b_{h_L }
\to e^{i2eVt} b_{h_L }$, which yields time dependent
operators $\cA_n(t)$. The  current is calculated  by second order
perturbation theory in $\cH^{tun-mag}$ yielding
\bea
I &=&\sum_n 2ne   X_n^{ret} (2eV)    \nonumber\\
X_n^{ret} (\omega) &=& i\int_{0}^\infty dt
e^{i\omega t} \langle \left[ A_n^\nd (t) ,A_n^\dagger\right]\rangle
\eea
For  singular contributions $I^{sing}$, we  ignore superconducting
condensate fluctuations $b^\dagger_h-\langle b^\dagger_h\rangle$, which have a
smooth spectrum. Similarly, we ignore the frequency dependence of
$T_n(\omega)$.  Setting
$b_R^\dagger \to \langle b^\dagger_h\rangle  $ and $b_L^\dagger \to e^{i2eVt}
\langle b^\dagger_h\rangle $  leads to 
\bea
 I^{sing} &=&\sum_n 2ne \sum_{|q_x|\le \pi/d,|q_y|\le \pi/W}     \langle b^\dagger_h\rangle^{4n}
|T_n[ \bq] |^2 \nonumber\\
&& \times \Im \sum_{\omega} S(\bq, i\omega+2neV+i0^+)  S(-\bq,
i\omega)
\eea
where the barrier dimensions are $d\times W$ (see Fig.\ref{fig:diag}), and
$\sum_\omega$ is a Matsubara sum.

\begin{figure}[htbp]
  \begin{center}
    \leavevmode \epsfxsize0.75\columnwidth
    \epsfbox{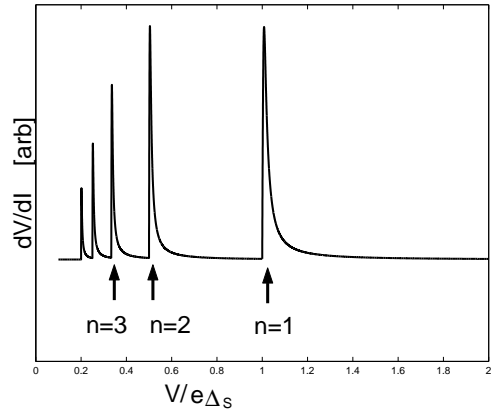}\vskip0.2pc
    \caption{ {\bf pSO(5)  Andreev peaks.}  Eq.
\ref{peaks} is plotted for a choice of $t_n/\Delta_s^{3/2}  = 2^{-n} 10^{-4}$,
$n\le 5$, and a  background conductance of unity. Below the $n$-th peak, the
excess tunneling charge is $2ne$, rather than BTK's $ne$.} \label{fig:peaks}
\end{center} \end{figure}

For a nearly antiferromagnetic ``N'' barrier,
$T_n(\bx-\bx')$ in (\ref{H-tun}) decays  slowly  with the distance
between magnons.
Thus for a narrow barrier $d<<W$,  the magnons are excited at
$q_y\approx 0$, and the momentum sum reduces to a one dimensional sum over
$q_x$. At zero temperature we obtain
\bea
I^{sing} &=&\sum_n 2ne     \langle b^\dagger_h\rangle^{4n}
|T_n[ 0] |^2 \nonumber\\  &&
\times  s_0^2 \int {dq_x\over 2\pi}  {\delta(2neV-2\sqrt{c^2 q_x^2
+\Delta_s^2}) \over 2 (\Delta_s^2 + c^2 q_x^2)  }\nonumber\\
&\approx & \sum_n t_n {\theta(neV-\Delta_s)  \over  \Delta_s^{3/2}
\sqrt{ neV-\Delta_s } }
\label{peaks}
\eea
The last expression emphasizes the singular form of
$I^{sing}(V,\Delta_s)$ at the peaks.
For a large background  conductance $dI/dV>> dI^{sing} /dV$, the
inverse square root singularities in $I^{sing}$ create
peaks in the differential resistance $dV/dI$ at voltages
\be
V_n=\Delta_s/(ne),~~~~    n=1,2,\ldots ,~~~ Q_n= 2ne
\ee
where $Q_n$ is the excess tunneling charge below  the $n$-th
peak.  Note that $Q_n$ changes in increments of $2e$.
The differential resistance peak series is
depicted in Fig. \ref{fig:peaks}, for weak broadening of the singularities and
an arbitrary set of coefficients $t_n$.
%%%%%%%%%%%%%%%%%%%%%%%%%%%%%%%%%%%
 %%%%%%%%%%%%%%%%%%%%%
{\em Discussion.} We have seen that magnon pair creation induces peaks in
the differential resistance which are similar {\em in appearance}  to the
Andreev peaks of the KBT mechanism.  The crucial difference is that  here
{\em the singular  dissipative process  does not involve Cooper pair
breaking}, but low energy antiferromagnetic  excitations.

In KBT theory for two identical
superconductors, the peaks appear at voltages
 $V^{KBT}_n = 2\Delta/(ne),~~~~    n=1,2,\ldots$ which are the upper threshold
for tunneling of charges $Q_n=ne$. Thus,  KBT allows both even and odd number
of electron charges  to participate in the multiple Andreev reflection
process, as depicted in Fig. \ref{fig:btk}.  These charges change in
increments of $e$ at each peak. Therefore a decisive discrimination between the
processes of Fig.\ref{fig:btk} and Fig.\ref{fig:diag}  would be measurements
of the excess tunneling charge increments  at the peaks. A feasible 
method would perhaps  be low temperature shot noise\cite{shot} $S$
which measures  the tunneling charges via the relation\cite{shot-theory}
$S=2Q_n I(V_n) $. We eagerly look forward to the results of such experiments.

{\em
Acknowledgements.} We thank G. Deutscher, G. Koren, A. Mizel, O. Nesher and E.
Polturak for useful discussions. Support from the Israel Science Foundation
and the Fund for Promotion of Research at Technion is acknowledged.

\end{document}